\documentclass[%
 reprint,
 superscriptaddress,
 amsmath,amssymb,
 aps,
 prl,
 floatfix,
]{revtex4-2}

\usepackage{graphicx, epstopdf}% Include figure files
\usepackage{dcolumn}% Align table columns on decimal point
\usepackage{bm}% bold math
\usepackage{gensymb}
\newcommand{\picWidth}{8.6cm}
\newcommand{\widePicWidth}{17.2cm}
\usepackage[colorlinks,linkcolor=blue,hyperindex,CJKbookmarks]{hyperref}
\usepackage{flushend}
\makeatletter
\begin{document}
%\preprint{APS/123-QED}
\title{
Traces of Electron-Phonon Coupling in One-Dimensional Cuprates
% Phonon-Mediated Attraction in 1D Cuprates
}
% Force line breaks with \\
\author{Ta Tang}
    \affiliation{Department of Applied Physics, Stanford University, California 94305, USA.}
    \affiliation{Stanford Institute for Materials and Energy Sciences, SLAC National Accelerator Laboratory, 2575 Sand Hill Road, Menlo Park, California 94025, USA.}
\author{Brian Moritz}
    \affiliation{Stanford Institute for Materials and Energy Sciences, SLAC National Accelerator Laboratory, 2575 Sand Hill Road, Menlo Park, California 94025, USA.}
    %\affiliation{Department of Physics and Astrophysics, University of North Dakota, Grand Forks, ND 58202, USA.}
\author{Cheng Peng}
    \affiliation{Stanford Institute for Materials and Energy Sciences, SLAC National Accelerator Laboratory, 2575 Sand Hill Road, Menlo Park, California 94025, USA.}
\author{Z. X. Shen}
    \affiliation{Department of Applied Physics, Stanford University, California 94305, USA.}
    \affiliation{Stanford Institute for Materials and Energy Sciences, SLAC National Accelerator Laboratory, 2575 Sand Hill Road, Menlo Park, California 94025, USA.}
    \affiliation{Department of Physics, Stanford University, Stanford CA 94305, USA.}
    \affiliation{Geballe Laboratory for Advanced Materials, Stanford University, Stanford, CA 94305, USA.}
\author{Thomas P. Devereaux}
    \affiliation{Stanford Institute for Materials and Energy Sciences, SLAC National Accelerator Laboratory, 2575 Sand Hill Road, Menlo Park, California 94025, USA.}
    \affiliation{Geballe Laboratory for Advanced Materials, Stanford University, Stanford, CA 94305, USA.}
    \affiliation{Department of Materials Science and Engineering, Stanford University, Stanford CA 94305, USA.}

\date{\today}

\begin{abstract}
    The appearance of certain spectral features in one-dimensional (1D) cuprate materials has been attributed to a strong, extended attractive coupling between electrons. Here, using time-dependent density matrix renormalization group methods on a Hubbard-extended Holstein model, we show that extended electron-phonon ({\it e-ph}) coupling presents an obvious choice to produce such an attractive interaction that reproduces the observed spectral features and doping dependence seen in angle-resolved photoemission experiments: diminished $3k_F$ spectral weight, prominent spectral intensity of a holon-folding branch, and the correct holon band width.  
    While extended {\it e-ph} coupling does not qualitatively alter the ground state of the 1D system compared to the Hubbard model, it quantitatively enhances the long-range superconducting correlations and suppresses spin correlations. 
    Such an extended {\it e-ph} interaction may be an important missing ingredient in describing the physics of the structurally similar two-dimensional high-temperature superconducting layered cuprates, which may tip the balance between intertwined orders in favor of uniform $d$-wave superconductivity. 
\end{abstract}

\pacs{Valid PACS appear here}

\maketitle

% \section{Introduction}
The origin of high-temperature superconductivity found in layered, quasi-two-dimensional (2D) cuprates remains a puzzle despite concerted, continuous investigations over the last few decades. 
From the perspective of numerical simulations, simplified models such as the Hubbard and $t$-$J$ Hamiltonians have been studied extensively, which have produced rich physics relevant to cuprates such as antiferromagnetism, stripes, and strange metal behavior\cite{dagottoRev1994, arovasHubbardModel2022,qinHubbadrModel2022}. 
However, evidence that these simplified models possess a uniform $d$-wave superconducting ground state remains elusive. Quasi-long-range superconductivity has only been reported on small width cylinders \cite{whitetJLadder1997,ehlersHubbardDmrg2017,leBlancSol2dHubbard2015,jiangSuperconductivityDopedHubbard2019c,chungPlaquettePair2020,jiangGroundStatePhase2020,jiangGroundstatePhaseDiagram2021,gongRobustDwaveSc2021,jiangPairingPropertiesModel2022,simonscollaborationonthemany-electronproblemAbsenceSuperconductivityPure2020}, with strong competition from coexisting charge order. Superconducting correlations decay exponentially on the hole doped side for wider clusters, indicating the superconductivity is absent for parameters thought to be relevant to hole doped cuprates. 

These findings indicate that the Hubbard model is incomplete, at least for describing the cuprates and high-temperature superconductivity. The inclusion of additional ingredients, such as phonons, which manifest as kinks or replica bands in photoemission measurements\cite{lanzaraEvidenceUbiquitousStrong2001b, cukReviewElectronPhonon2005a, leeInterplayElectronLattice2006, heRapidChangeSuperconductivity2018}, may provide the crucial remedy. However, exact numerical simulations of the 2D Hubbard model already are challenging (the density matrix renormalization group (DMRG) method is limited by the growth of entanglement entropy and determinant quantum Monte Carlo (DQMC) and related methods suffer from the fermion sign problem); and adding bosonic degrees of freedom creates an even more daunting problem. The task may be made easier, with more numerical control, by turning to the simpler, yet structurally similar, one-dimensional (1D) cuprates.
Recent angle-resolved photoemission spectroscopy (ARPES) experiments on the 1D cuprate $\mathrm{Ba}_{2-x}\mathrm{Sr}_x\mathrm{CuO}_{3+\delta}$~\cite{chenAnomalouslyStrongNearneighbor2021} provide an excellent platform for testing theoretical models. Modeling in 1D has both well-established theory and numerical simulations can be performed with a higher degree of control and accuracy.  
The measured single-particle spectra provide a detailed proving ground for assessing the impact of terms added to model Hamiltonians.
Reference~\cite{chenAnomalouslyStrongNearneighbor2021} showed that the simple Hubbard model fails to reproduce salient details of the spectra near the Fermi surface: a prominent holon-folding (hf)-branch emanates from $k_F$ and quickly fades away with doping. This spectral feature, and its doping dependence, can be well reproduced when one includes a strong nearest-neighbor attractive interaction $V \sim -t$ in the model Hamiltonian. A natural near-neighbor attraction exists in the Hubbard model, evident when downfolding to the $t$-$J$ model, but such a weak attraction ($\sim -J/4$) cannot account for the observed effect. Rather, this strong attraction likely originates from extended electron-phonon ({\it e-ph}) coupling, as discussed in recent work~\cite{chenAnomalouslyStrongNearneighbor2021,wangPhononMediatedLongRangeAttractive2021}. 

%Rather than working with an effective attractive interaction, 
To investigate the influence of the extended {\it e-ph} coupling, %directly work with the underlying electron-phonon Hamiltonian. Specifically, 
in this paper a time-dependent density matrix renormalization group (tDMRG) method is employed to study the single-particle spectral function and ground state properties of a 1D Hubbard-extended Holstein model. 
The extended {\it e-ph} coupling quantitatively reproduces the dominant hf-branch seen in experiments, while also correctly reproducing the holon branch band width, matching the observed spectra. Approximating this model using an effective nearest-neighbor attraction V fails to reproduce all of these features. Moreover, while the extended {\it e-ph} coupling does not qualitatively alter the ground state obtained from the Hubbard model, which qualitatively remains a Luttinger liquid with subdominant superconducting pair-field correlations that decay as a power law with distance, the results show that the extended {\it e-ph} coupling quantitatively enhances the superconducting pair-field correlations by reducing the overall exponent, making them longer-ranged. It is surmised that in two dimensions an extended {\it e-ph} coupling may tip the balance between different phases and help to realize a dominant $d$-wave superconducting ground state.

\begin{figure}[tb!]
    \begin{center}
        \includegraphics[width=\picWidth]{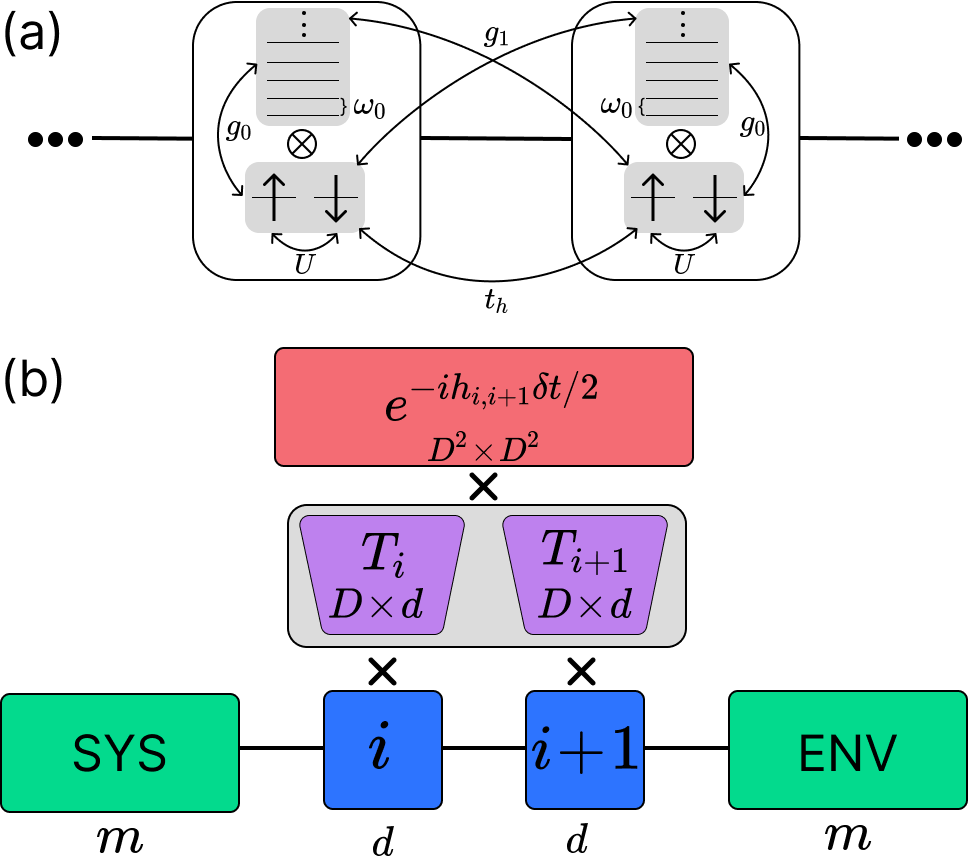}
    \end{center}
    \caption{
        (a) Schematic for the one dimensional Hubbard-extended Holstein model. On each site, the local Hilbert space is a direct product of phonon and charge degrees of freedom. The charges of opposite spin interact with an on-site repulsion $U$ and can hop to neighboring sites. Local phonons with a frequency $\omega_0$ couple to both on-site and nearest-neighbor charges. (b) Schematic for the dynamical LBO. We keep the dimension of the effective Hilbert space of the system and environment blocks as $m$, respectively. Each site $i$ has $d$ optimized basis. The wave function is transformed to a $D \gg d$ bare basis ($D= D_{ch}\times D_{ph}$, where $D_{ch}=4$ represents the local charge Hilbert space dimension, and $D_{ph}$ is the bare phonon basis dimension) through a $D \times d$ transformation matrix, {\it i.e.} $T_i$, before applying the time evolution gate of shape $D^2 \times D^2$. Subsequently, a new optimal basis and transformation $T_i$ are obtained; and the wave function is projected to the new optimal basis before moving on to the next gate. 
    }
    \label{pic:diagram}
\end{figure}

\section{Models}
To produce an effective nearest-neighbor attractive interaction for charge, we consider an optical phonon mode, which couples to charge density beyond the local site. Previous estimates~\cite{wangPhononMediatedLongRangeAttractive2021} have shown that this Hubbard-extended Holstein model can produce an effective interaction on par with that extracted from ARPES experiments~\cite{chenAnomalouslyStrongNearneighbor2021} for a reasonable phonon frequency and {\it e-ph} coupling strength. For simplicity and to achieve better numerical convergence, here, we consider only on-site and nearest-neighbor {\it e-ph} coupling (see Fig.~\ref{pic:diagram} (a)). This Hubbard-extended Holstein Hamiltonian takes the form
\begin{eqnarray}
    H &=& H_{el} + \omega_0\sum_i \hat{a}^\dagger_i\hat{a}_i \nonumber \\
    &&\quad + g_0\sum_i \hat{n}_i (\hat{a}^\dagger_i + \hat{a}_i) + g_1\sum_{\left<ij\right>} \hat{n}_i (\hat{a}^\dagger_j + \hat{a}_j),
    \label{eq:hhm}
\end{eqnarray}
where $\hat{a}^\dagger_i$ and $\hat{a}_i$ are the phonon ladder operators on site $i$, $\hat{n}_i$ is the total charge number operator on site $i$, $\omega_0$ is the phonon frequency, $g_0$ is the on-site {\it e-ph} coupling, $g_1$ is the nearest-neighbor {\it e-ph} coupling, and $\left<ij\right>$ sums over nearest-neighbors.
$H_{el}$ denotes the electronic part of the Hamiltonian, a 1D single-band Hubbard model,
\begin{equation}
H_{el} = -t_h\sum_{\left<ij\right>\sigma}(\hat{c}^\dagger_{i\sigma}\hat{c}_{j\sigma} + h.c.) + U\sum_{i}\hat{n}_{i\uparrow}\hat{n}_{i\downarrow},
\label{eq:hubbard}
\end{equation}
where $\hat{c}^\dagger_{i\sigma}$ ($\hat{c}_{i\sigma}$) is the charge creation (annihilation) operator on site $i$ for spin $\sigma$, $\hat{n}_{i\sigma}$ is the charge number operator on site $i$ for spin $\sigma$, and $U$ is the on-site repulsion. To avoid confusion with the time variable $t$, we use $t_h$ to denote the hopping integral. For comparison, we also evaluate the extended-Hubbard model, which introduces a nearest-neighbor attractive interaction,
\begin{equation}
H_{v} = H_{el} + V\sum_{\left<ij\right>}\hat{n}_i\hat{n}_j,
\label{eq:extended-hubbard}
\end{equation}
where $\hat{n}_i$ and $\hat{n}_j$ are total charge number operators on neighboring sites.

Unless otherwise specified, we use the following parameters in our simulations: $U=8t_h$, $\omega_0=0.2t_h$, $g_0=0.3t_h$, $g_1=0.15t_h$ and $V=-t_h$. The values chosen for $U$ and $V$ were those that produced the best fit of the ARPES experimental spectra using cluster perturbation theory (CPT)~\cite{senechalSpectralWeightHubbard2000a, senechalCPT2002} for an effective extended-Hubbard model~\cite{chenAnomalouslyStrongNearneighbor2021}; and the {\it e-ph} couplings $g_0$ and $g_1$ fall within the range estimated in Ref.~\cite{wangPhononMediatedLongRangeAttractive2021}. Here, we use a larger phonon frequency than that used in Ref.~\cite{wangPhononMediatedLongRangeAttractive2021} for better numerical convergence, but expect that a smaller phonon frequency would produce a stronger effective attraction, which would further enhancing the hf-branch; although, one would need to ensure that the stronger effective coupling would not lead to phase separation.   
\begin{figure}[tb!]
    \begin{center}
        \includegraphics[width=\picWidth]{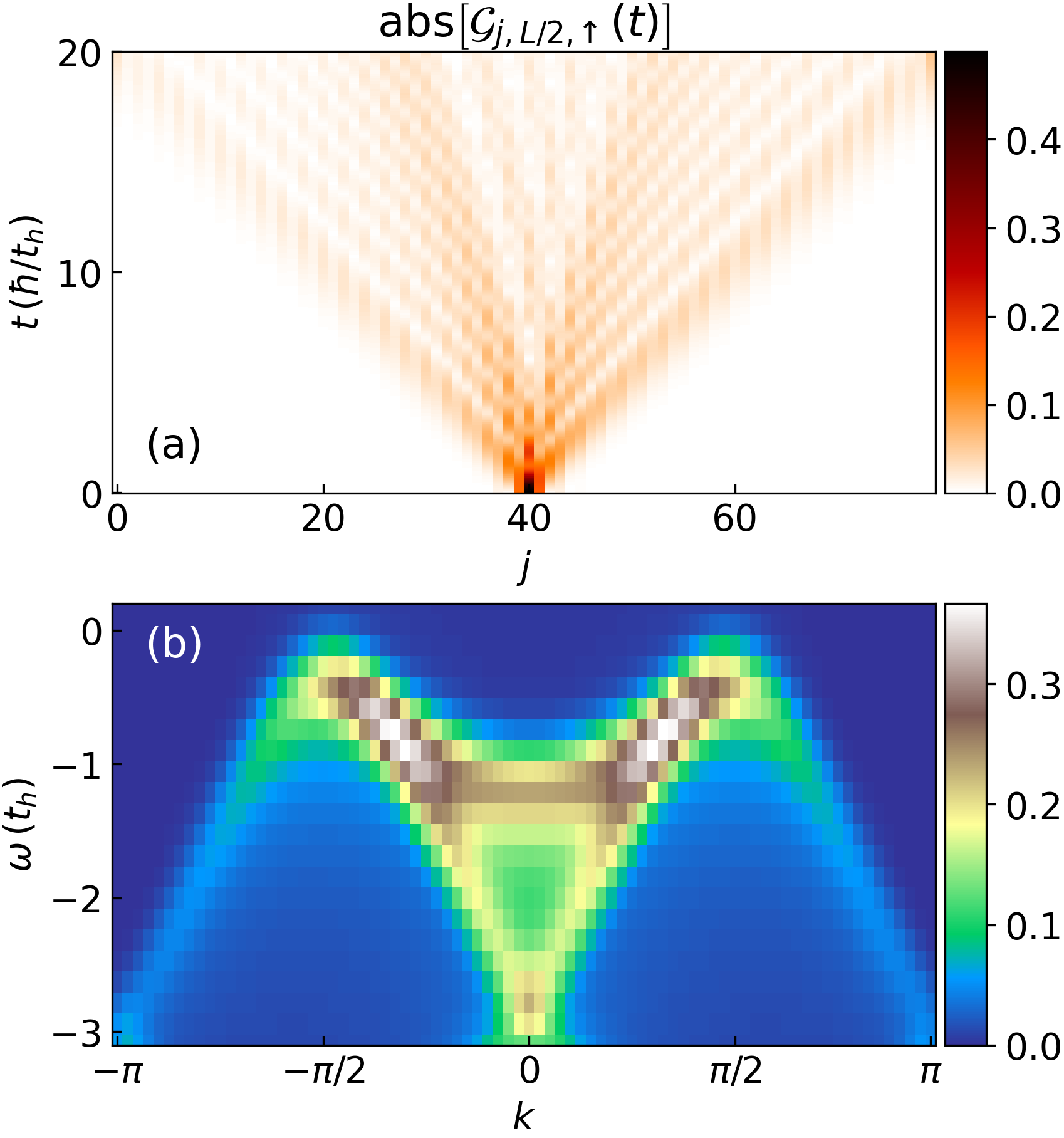}
    \end{center}
    \caption{
        (a) The lesser Green's function $\mathcal{G}^{<}_{j,L/2,\uparrow}(t)$ for an 80-site chain at half-filling for the Hubbard model. Time is measured in units of $\hbar/t_h$ and $\hbar = 1$ in our calculation. We use a time step $\delta t = 0.04 t_h^{-1}$ and evolve the system for a total time $T=20 t_h^{-1}$. (b) The single-particle spectral function obtained by Fourier transform of $\mathcal{G}^{<}$ in (a), with energy and momentum broadening of $\sigma_{\omega} = 0.2t_{h}$ and $\sigma_{k} = 2\pi/L$, respectively. 
    }
    \label{pic:akw_example}
\end{figure}

\begin{figure*}[htpb!]
    \begin{center}
        \includegraphics[width=\widePicWidth]{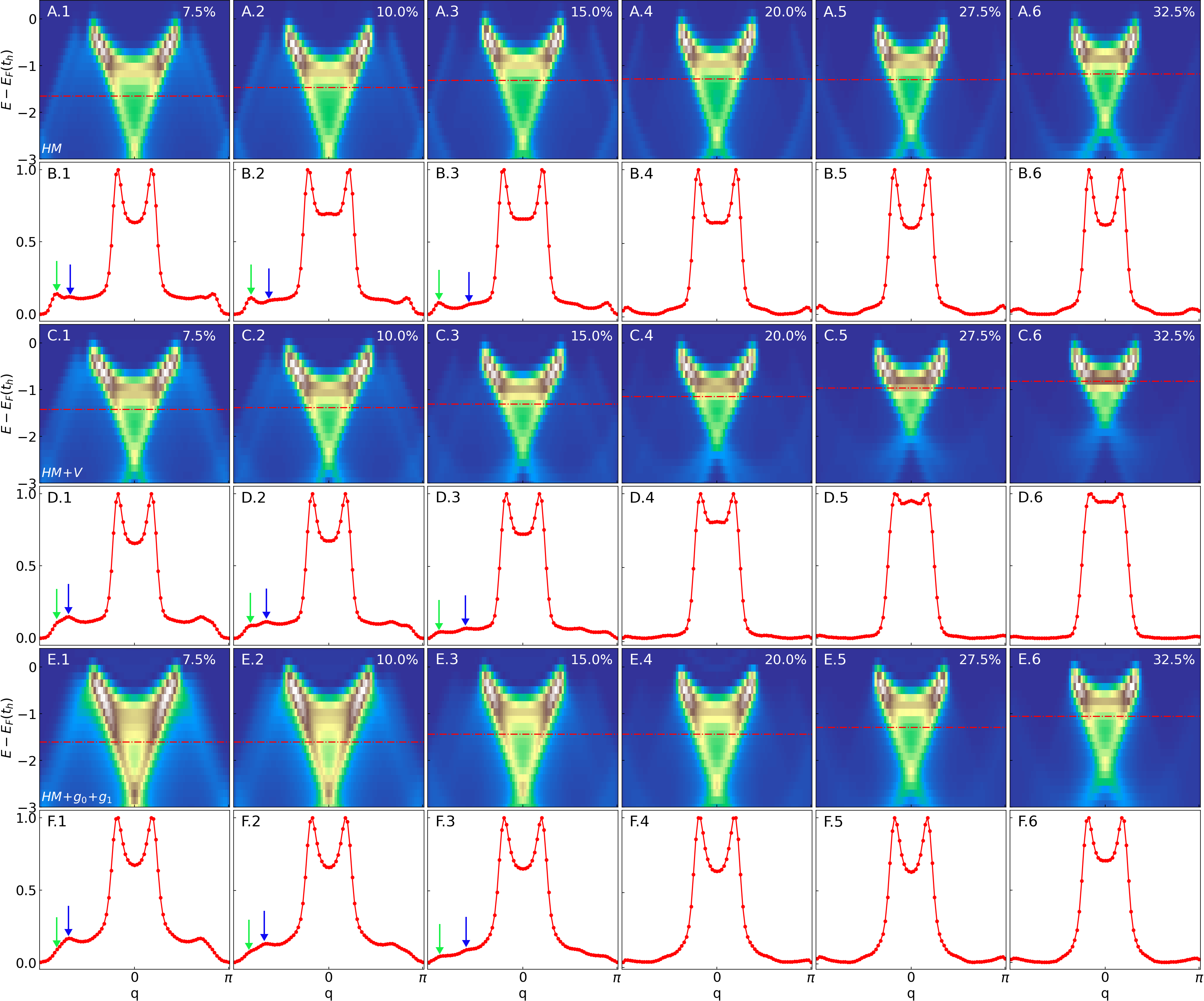} 
    \end{center}
    \caption{
        Single-particle spectra for (A) the Hubbard model ($HM$), (C) the extended Hubbard model ($HM+V$), and (E) the Hubbard-extended Holstein model ($HM+g_0+g_1$), with increasing doping from column 1 to 6. 
        Panels (B), (D), and (F) show representative momentum distribution curves (MDCs), corresponding to the cuts given by the red dashed line for each of the spectra in (A), (C), and (E), respectively. The MDCs are chosen $\sim t_{h}$ above the bottom of the holon branch to ensure that the main holon peaks are at roughly the same position for different dopings and for different models, providing equivalent MDCs for comparison. The green and blue arrows mark the positions of $3k_F$ and hf branches, respectively.
        One can clearly see that at lower doping($<20\%$), adding nearest neighbor attraction $V$ or extended {\it e-ph} coupling can enhance the hf branch while suppress the $3k_F$ branch. Above $20\%$ doping, both peaks fade away quickly. Here, the energy and momentum broadening of the spectra are $\sigma_\omega=0.18t_h$ and $\sigma_k=2\pi/L$. 
    }
    \label{pic:akw}
\end{figure*}

We use DMRG~\cite{whiteDensityMatrixFormulation1992, whiteDMRG1993} to obtain the ground states of the models defined in Eqs.\,\ref{eq:hhm}, \ref{eq:hubbard}, and \ref{eq:extended-hubbard}; and we use tDMRG~\cite{vidalEfficientSimulationOneDimensional2004, whiteRealTimeEvolutionUsing2004, paeckelTimeevolutionMethodsMatrixproduct2019} to obtain real-frequency spectra from the Fourier transform of time-dependent correlators of the form $\left<\hat{O}^\dagger_i(t)\hat{O}_j(0)\right>$.  
To efficiently deal with the infinite phonon Hilbert space on each site, 
we adopt a local basis optimization (LBO) for the ground state~\cite{zhangDensityMatrixApproach1998a} and a dynamical LBO for time evolution~\cite{brocktMatrixproductstateMethodDynamical2015a}, as schematically shown in Fig.~\ref{pic:diagram}(b). 
Details about the method and numerical simulation are provided in the supplementary material.

\section{Single Particle Spectral Function}
Fig.~\ref{pic:akw_example} displays the lesser Green's function $\mathcal{G}^{<}_{j,L/2,\uparrow}(t)$, defined as $\mathcal{G}^{<}_{mn\sigma}(t) = i\left<\hat{c}^\dagger_{m\sigma}(t) \hat{c}_{n\sigma}(0)\right>$, and the corresponding single-particle removal spectra, obtained for the Hubbard model on an $80$-site chain at half-filling. In Fig.~\ref{pic:akw_example}(a), following the removal of an electron from the center of the chain, one can see that the propagator attains a significant value at the two chain ends within a time $T \sim 20 t_h^{-1}$, which sets the maximum real-time propagation for the simulation. Padding the Green's function with zeros from time $T$ to time $2T$ limits the frequency resolution of a fast Fourier transform (FFT) to $\omega_{n+1}-\omega_n = \pi/T \sim 0.16\,t_{h}$. This provides a rather coarse resolution, but it is nevertheless more than adequate for comparison to the experimental ARPES spectra from the 1D chain cuprate, which is rather broad~\cite{chenAnomalouslyStrongNearneighbor2021}. 
The single-particle spectrum, which is obtained using the tDMRG method and shown in Fig.~\ref{pic:akw_example}(b), agrees well with the results from cluster perturbation theory~\cite{senechalSpectralWeightHubbard2000a, senechalCPT2002, chenAnomalouslyStrongNearneighbor2021}, dynamical DMRG, and the Bethe ansatz~\cite{jeckelmannDynamicalDensitymatrixRenormalizationgroup2002, benthien1dHubbardDDMRG2004, essler1dHubbard2005, kohnoSpectralPropertiesMott2010}. 
There are clear spinon and holon branches, demonstrating spin-charge separation in 1D. In the following, we use a chain of length $L=80$ to compute and compare the single particle spectral function of different models. A small broadening is used to give the spectra a high resolution, at least when compared with the experiment data, to better observe how different models affect the salient spectral features.

Fig.~\ref{pic:akw}\,(A.1-6) show the single-particle removal spectra of the Hubbard model across a range of doping. As observed in experiment, splitting between the spinon and holon branches persists with doping. Our results correspond well to previous Hubbard model results on 1D and quasi-1D systems from dynamical DMRG and the Bethe ansatz~\cite{benthien1dHubbardDDMRG2004, kohnoSpectralPropertiesMott2010}, and also are consistent with spectra near the Fermi level from determinant quantum Monte Carlo and DMRG calculations of the multi-band Hubbard model, which includes oxygen $p$-orbitals~\cite{liParticleholeAsymmetryDynamical2021}. Here, we will focus on two spectral features: the branch of the removal spectrum emanating from $k_F$, which disperses downward toward $\pi$, hereafter the hf-branch; and the $3k_F$-branch (or more precisely $2\pi-3k_F$), which also disperses downward toward $\pi$, but from $3k_F$. In the MDCs obtained from Hubbard model (Figs.~\ref{pic:akw}(B.1-6)), one sees that between these two features the $3k_F$-peak is dominant. 
This result is contradictory to experimental observations, where the hf-peak is dominant and the $3k_F$-peak is barely visible ~\cite{chenAnomalouslyStrongNearneighbor2021}.

\begin{figure}[tb!]
    \begin{center}
        \includegraphics[width=\picWidth]{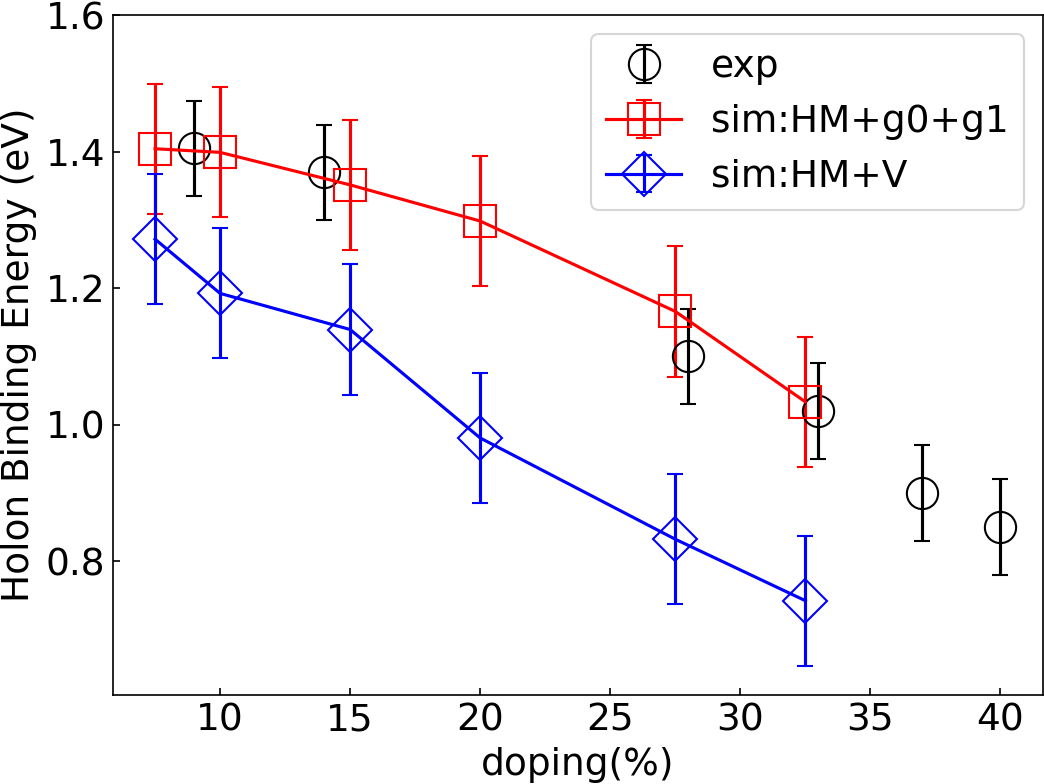}
    \end{center}
    \caption{
        Comparison of experimental and simulated holon binding energy at momentum $k=0$. The experiment data (open circle) are taken from Ref.\cite{chenAnomalouslyStrongNearneighbor2021}. For the holon binding energy, the Hubbard-extended Holstein model (open square) matches the experiment data very well, while the extended-Hubbard model (open diamond) deviates from experiment at higher doping. Here we take $t_h=530$meV.
    }
    \label{pic:bandwidth}
\end{figure}

\begin{figure*}[htpb!]
    \begin{center}
        \includegraphics[width=\widePicWidth]{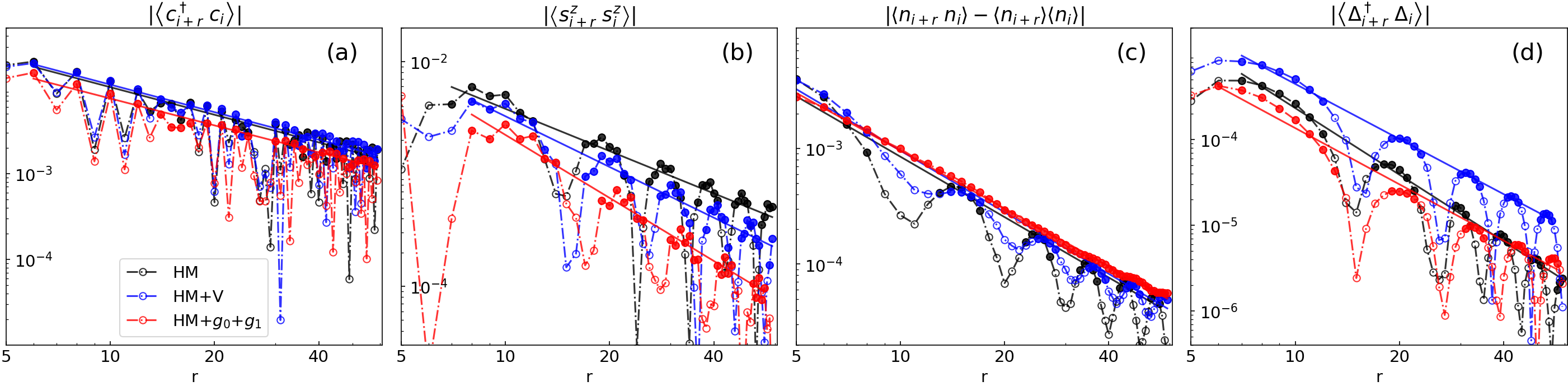} 
    \end{center}
    \caption{
        Correlation functions for different models. 
        Correlators for the Hubbard (black) and extended Hubbard (blue) models are plotted for comparison. 
        The Hubbard-extended Holstein model (red) results were obtained for $\omega_0=0.2t_h$, $g_0=0.3t_h$ and $g_1=0.15t_h$. 
        The straight line fits follow a power law decay $\sim r^{-K}$ to extract effective Luttinger exponents for different correlation functions. Filled circles show data used for fitting \cite{jiangGroundStatePhase2020}.
    }
    \label{pic:corr}
\end{figure*}

In Figs.~\ref{pic:akw}(C.1-6 and D.1-6), we confirm  that adding a nearest-neighbor attractive interaction $V=-t_h$ enhances the hf-branch and produces spectra that are visibly more consistent with the experimental data at lower doping~\cite{chenAnomalouslyStrongNearneighbor2021}.
As we mentioned previously, this attractive interaction likely originates from {\it e-ph} coupling. Here, we also simulate the underlying {\it e-ph} Hamiltonian, with the results shown in Figs.~\ref{pic:akw}(E.1-6). 
Below $20\%$ doping, one sees an enhanced hf-branch, while the $3k_F$-branch has been suppressed significantly by the {\it e-ph} coupling (see Figs.~\ref{pic:akw}(E.1-3 and F.1-3)). In all three models, the intensities in both the hf- and $3k_{F}$-branches become barely perceptible beyond $\sim 20\%$ doping. 
Using a larger broadening to compare more closely with the experimental spectra and to extract intensities by fitting MDCs results in a doping-dependent intensity of hf-peak that matches well to the analyzed ARPES data (see Figs.~\ref{pic:akw_phonon_broad} and \ref{pic:phonon_hf_intensity} in the Supplementary Material).

One significant difference between spectra for the extended Hubbard model and the Hubbard-extended Holstein model is that the nearest-neighbor attractive interaction in the extended Hubbard model significantly shrinks the holon bandwidth at higher doping (see Figs.~\ref{pic:akw}(C.1-6)). In Fig.~\ref{pic:bandwidth}, we plot the holon binding energy at $k=0$ as a function of doping to reflect the change of the holon bandwidth. By comparison, one sees that the results from the Hubbard-extended Holstein model are more consistent with the ARPES data, as the {\it e-ph} interaction would renormalize the holon-branch only within $\sim \omega_0$ of the Fermi energy.

\section{Ground state correlation functions}
The good agreement with ARPES measurements begs the question: How does the extended {\it e-ph} interaction affect the ground state? 
As a first step towards understanding this question, we study the ground state correlation functions of the 1D Hubbard-extended Holstein model (as well as the Hubbard and extended Hubbard models) at $10\%$ hole doping using a $120$-site chain to observe relatively long-distance behavior. 
We measure equal-time correlation functions of the form $\left<\hat{O}_{i+r}\hat{O}_i\right>$, averaged over $5$ reference points ({\it i.e.} $i=L/4-1, L/4, ..., L/4+3$) for each $r$, where $r$ is the distance between two sites along the chain between $0$ to $L/2$. In this way, the measurements fall roughly within the center half of the chain to reduce boundary effects. 

Our results suggest that the ground state of the Hubbard-extended Holstein model in 1D is consistent with a Luttinger liquid (LL)~\cite{giamarchiQuantumPhysicsOne2003}, as evidenced by the slow decay of the single-particle Green's function defined as $G_{\sigma}(r) = \langle \hat{c}^{\dagger}_{i+r,\sigma} \hat{c}_{i,\sigma}\rangle$. Specifically, $G_\sigma(r)$ as shown in Fig.~\ref{pic:corr}(a) can be very well fitted by a power law, {\it i.e.} $G_{\sigma}(r)\sim r^{-K_G}$. The decaying behaviour of the single-particle Green function for each of the three different models is qualitatively consistent with the Luttinger exponent $K_G \sim 1$. We provide the value of $K_G$ extracted from each model in Table~\ref{tab:exponents}. For completeness, we have calculated the spin-spin correlation function defined as $F(r) = \langle \mathbf{S}_{i+r}\cdot \mathbf{S}_{i}\rangle$. As shown in Fig.~\ref{pic:corr}(b), $F(r)$ also appears to decay as a power law, $F(r)\sim r^{-K_s}$, but with a larger exponent than the single-particle correlation, $K_s>K_G$, also consistent LL behavior. Note that the extended {\it e-ph} interaction produces a larger suppression of the spin-spin correlations, resulting in the largest $K_s$ among the three models. The charge density-density fluctuation correlations, defined as $D(r) = \langle \hat{n}_{i+r}\hat{n}_{i}\rangle-\langle \hat{n}_{i+r} \rangle \langle \hat{n}_{i} \rangle$, also appear quasi-long-ranged with a Luttinger exponent $K_c$, also shown in Table~\ref{tab:exponents}. 

The most intriguing aspect of the extended interactions may be their influence on superconductivity, tested through the equal-time spin-singlet superconducting pair-field correlation function, $P(r) = \langle \Delta^{\dagger}_{i+r} \Delta_{i}\rangle$, where $\Delta_{i} = \frac{1}{\sqrt{2}}(\hat{c}_{i\uparrow}\hat{c}_{i+1,\downarrow}-\hat{c}_{i\downarrow}\hat{c}_{i+1,\uparrow})$ is the spin-singlet pair-field annihilation operator. As expected for a LL, $P(r)$ decays as a power law, with $K_{sc}>2$ for all three models, as shown in Fig.~\ref{pic:corr}(d) and Table~\ref{tab:exponents}. Most importantly, not only does the nearest-neighbor attractive interaction enhance $P(r)$, but the extended {\it e-ph} coupling also produces a noticeably smaller $K_{sc}$ compared to the Hubbard model alone. Taken together, while the extended {\it e-ph} interaction itself does not qualitatively alter the ground state of the system in 1D, it does quantitatively enhance the strength of singlet superconducting pair-field correlations and suppress spin-spin correlations.

\section{Discussion}
In summary, the inclusion of extended electron-lattice couplings is crucially important for reproducing many of the observed spectral features in ARPES. The extended {\it e-ph} coupling reproduces well the intensity and doping dependence of the hf-feature, the reduced $3k_F$-feature, and gives the right doping dependence of the holon band width.  
As more experimental results emerge for doped 1D systems, it would be beneficial to check the impact of {\it e-ph} coupling on other measurements, such as the dynamical spin structure factor and phonon dispersion. 
%However, we note that the relevant energy scale would be smaller ($\sim J, \sim \omega_0$) compared to the single particle spectral function, therefore to obtain good energy resolution likely would require a longer time evolution using tDMRG. 

Our results show that the ground state of the 1D Hubbard-extended Holstein model remains a Luttinger liquid with a single-particle correlation exponent $K_G\sim 1$ and subdominant superconducting correlations. 
However, quantitatively, the extended {\it e-ph} coupling helps to suppress the spin correlations $F(r)$, while simultaneously enhancing the superconducting pair-field correlations $P(r)$. 
Importantly, while the inclusion of a simple effective nearest-neighbor attractive interaction to approximate the extended {\it e-ph} coupling can produce a similar enhancement of the hf-branch, it fails to produce the right holon band width. It also enhances $P(r)$ and gives an $K_{sc}$ close to the one produced by the extended {\it e-ph} coupling, but overestimates the magnitude of $P(r)$ and does not suppress $F(r)$ as effectively as the extended {\it e-ph} coupling. 

It is of course an open and interesting question to determine whether the agreement between numerical results for the Hubbard-extended Holstein model and ARPES translates to dimensions greater than 1. A recent DMRG study on 4-leg ladders has shown that an effective nearest-neighbor electron-electron attraction can result in dominant quasi-long-range $d$-wave superconducting correlations, where the crossover between dominant superconducting and CDW correlations occurs near $V \sim -t_h$~\cite{pengEnhancedSuperconductivityNearneighbor2022}. Yet that ground state remains qualitatively consistent with a Luther-Emery liquid, as found in the simple Hubbard model on the same ladder. As it appears that power-law decay of superconducting correlations cede to a short-range exponential decay of correlations as the hole-doped ladder system goes to 2D, a boost of superconducting pairing from extended electron-lattice coupling could be pivotal to both qualitatively and quantitatively change the nature of the ground state. While this remains a topic of investigation, our results encourage additional study on the influence of phonon degrees of freedom in 2D models, which finally may help to realize a $d$-wave superconducting ground state. 

%% exponents
\begin{table}[htp!]
    \begin{tabular}{p{3cm}<{\centering} p{1.2cm}<{\centering} p{1.2cm}<{\centering} p{1.2cm}<{\centering} p{1.2cm}<{\centering}}
        \hline
        Model              & $K_G$ & $K_s$ & $K_c$ & $K_{sc}$ \\
        \hline
        $HM$            & 1.07(6)  & 1.25(5)      & 1.74(5)    & 2.58(4)          \\
        $HM+V$          & 1.07(5)  & 1.50(7)      & 1.77(4)    & 2.14(3)          \\
        $HM+g_0+g_1$    & 1.04(4)  & 1.87(6)      & 1.66(1)    & 2.18(6)          \\
        \hline
    \end{tabular}
    \caption{Comparison of exponents extracted for various correlation functions for different models. The exponents $K$ are extracted from the fits $\sim r^{-K}$ shown in Fig.~\ref{pic:corr}.}
    \label{tab:exponents}
\end{table}

\section*{Acknowledgement}
The authors would like to thank Hongchen Jiang, Yao Wang and Zhuoyu Chen for helpful discussions and suggestions. This work was supported by the U.S. Department of Energy, Office of Basic Energy Sciences, Division of Materials Sciences and Engineering, under Contract No.~DE-AC02-76SF00515. The computational results utilized the resources of the National Energy Research Scientific Computing Center (NERSC) supported by the U.S. Department of Energy, Office of Science, under Contract No.~DE-AC02-05CH11231. Some of the computing for this project was performed on the Sherlock cluster. We would like to thank Stanford University and the Stanford Research Computing Center for providing computational resources and support that contributed to these research results.

\bibliography{dmrg}

\begin{thebibliography}{10}

\bibitem{dagottoRev1994}
Elbio Dagotto.
\newblock Correlated electrons in high-temperature superconductors.
\newblock {\em Rev. Mod. Phys.}, 66:763--840, Jul 1994.

\bibitem{arovasHubbardModel2022}
Daniel~P. Arovas, Erez Berg, Steven~A. Kivelson, and Srinivas Raghu.
\newblock The hubbard model.
\newblock {\em Annual Review of Condensed Matter Physics}, 13(1):239--274,
  2022.

\bibitem{qinHubbadrModel2022}
Mingpu Qin, Thomas Sch\"{a}fer, Sabine Andergassen, Philippe Corboz, and
  Emanuel Gull.
\newblock The hubbard model: A computational perspective.
\newblock {\em Annual Review of Condensed Matter Physics}, 13(1):275--302,
  2022.

\bibitem{whitetJLadder1997}
Steven~R. White and D.~J. Scalapino.
\newblock Ground states of the doped four-leg t-j ladder.
\newblock {\em Phys. Rev. B}, 55:R14701--R14704, Jun 1997.

\bibitem{ehlersHubbardDmrg2017}
G.~Ehlers, S.~R. White, and R.~M. Noack.
\newblock Hybrid-space density matrix renormalization group study of the doped
  two-dimensional hubbard model.
\newblock {\em Phys. Rev. B}, 95:125125, Mar 2017.

\bibitem{leBlancSol2dHubbard2015}
J.~P.~F. LeBlanc, Andrey~E. Antipov, Federico Becca, Ireneusz~W. Bulik, Garnet
  Kin-Lic Chan, Chia-Min Chung, Youjin Deng, Michel Ferrero, Thomas~M.
  Henderson, Carlos~A. Jim\'enez-Hoyos, E.~Kozik, Xuan-Wen Liu, Andrew~J.
  Millis, N.~V. Prokof'ev, Mingpu Qin, Gustavo~E. Scuseria, Hao Shi, B.~V.
  Svistunov, Luca~F. Tocchio, I.~S. Tupitsyn, Steven~R. White, Shiwei Zhang,
  Bo-Xiao Zheng, Zhenyue Zhu, and Emanuel Gull.
\newblock Solutions of the two-dimensional hubbard model: Benchmarks and
  results from a wide range of numerical algorithms.
\newblock {\em Phys. Rev. X}, 5:041041, Dec 2015.

\bibitem{jiangSuperconductivityDopedHubbard2019c}
Hong-Chen Jiang and Thomas~P. Devereaux.
\newblock Superconductivity in the doped {{Hubbard}} model and its interplay
  with next-nearest hopping t'.
\newblock {\em Science (New York, N.Y.)}, 365(6460):1424--1428, 2019.

\bibitem{chungPlaquettePair2020}
Chia-Min Chung, Mingpu Qin, Shiwei Zhang, Ulrich Schollw\"ock, and Steven~R.
  White.
\newblock Plaquette versus ordinary $d$-wave pairing in the
  ${t}^{\ensuremath{'}}$-hubbard model on a width-4 cylinder.
\newblock {\em Phys. Rev. B}, 102:041106, Jul 2020.

\bibitem{jiangGroundStatePhase2020}
Yi-Fan Jiang, Jan Zaanen, Thomas~P. Devereaux, and Hong-Chen Jiang.
\newblock Ground state phase diagram of the doped {{Hubbard}} model on the
  four-leg cylinder.
\newblock {\em Physical Review Research}, 2(3):033073, July 2020.

\bibitem{jiangGroundstatePhaseDiagram2021}
Shengtao Jiang, Douglas~J. Scalapino, and Steven~R. White.
\newblock Ground-state phase diagram of the t-t{${'}$}-{{J}} model.
\newblock {\em Proceedings of the National Academy of Sciences},
  118(44):e2109978118, November 2021.

\bibitem{gongRobustDwaveSc2021}
Shoushu Gong, W.~Zhu, and D.~N. Sheng.
\newblock Robust $d$-wave superconductivity in the square-lattice
  $t\text{\ensuremath{-}}j$ model.
\newblock {\em Phys. Rev. Lett.}, 127:097003, Aug 2021.

\bibitem{jiangPairingPropertiesModel2022}
Shengtao Jiang, Douglas~J. Scalapino, and Steven~R. White.
\newblock Pairing properties of the $t$-$t'$-$t''$-$j$ model, June 2022.

\bibitem{simonscollaborationonthemany-electronproblemAbsenceSuperconductivityPure2020}
{Simons Collaboration on the Many-Electron Problem}, Mingpu Qin, Chia-Min
  Chung, Hao Shi, Ettore Vitali, Claudius Hubig, Ulrich Schollw{\"o}ck,
  Steven~R. White, and Shiwei Zhang.
\newblock Absence of {{Superconductivity}} in the {{Pure Two-Dimensional
  Hubbard Model}}.
\newblock {\em Physical Review X}, 10(3):031016, July 2020.

\bibitem{lanzaraEvidenceUbiquitousStrong2001b}
A.~Lanzara, P.~V. Bogdanov, X.~J. Zhou, S.~A. Kellar, D.~L. Feng, E.~D. Lu,
  T.~Yoshida, H.~Eisaki, A.~Fujimori, K.~Kishio, J.-I. Shimoyama, T.~Noda,
  S.~Uchida, Z.~Hussain, and Z.-X. Shen.
\newblock Evidence for ubiquitous strong electron\textendash phonon coupling in
  high-temperature superconductors.
\newblock {\em Nature}, 412(6846):510--514, August 2001.

\bibitem{cukReviewElectronPhonon2005a}
T.~Cuk, D.~H. Lu, X.~J. Zhou, Z.-X. Shen, T.~P. Devereaux, and N.~Nagaosa.
\newblock A review of electron\textendash phonon coupling seen in the
  high-{{Tc}} superconductors by angle-resolved photoemission studies
  ({{ARPES}}).
\newblock {\em physica status solidi (b)}, 242(1):11--29, 2005.

\bibitem{leeInterplayElectronLattice2006}
Jinho Lee, K.~Fujita, K.~McElroy, J.~A. Slezak, M.~Wang, Y.~Aiura, H.~Bando,
  M.~Ishikado, T.~Masui, J.-X. Zhu, A.~V. Balatsky, H.~Eisaki, S.~Uchida, and
  J.~C. Davis.
\newblock Interplay of electron\textendash lattice interactions and
  superconductivity in {{Bi2Sr2CaCu2O8}}+{$\delta$}.
\newblock {\em Nature}, 442(7102):546--550, August 2006.

\bibitem{heRapidChangeSuperconductivity2018}
Y.~He, M.~Hashimoto, D.~Song, S.-D. Chen, J.~He, I.~M. Vishik, B.~Moritz, D.-H.
  Lee, N.~Nagaosa, J.~Zaanen, T.~P. Devereaux, Y.~Yoshida, H.~Eisaki, D.~H. Lu,
  and Z.-X. Shen.
\newblock Rapid change of superconductivity and electron-phonon coupling
  through critical doping in {{Bi-2212}}.
\newblock {\em Science}, 362(6410):62--65, October 2018.

\bibitem{chenAnomalouslyStrongNearneighbor2021}
Zhuoyu Chen, Yao Wang, Slavko~N. Rebec, Tao Jia, Makoto Hashimoto, Donghui Lu,
  Brian Moritz, Robert~G. Moore, Thomas~P. Devereaux, and Zhi-Xun Shen.
\newblock Anomalously strong near-neighbor attraction in doped {{1D}} cuprate
  chains.
\newblock {\em Science}, September 2021.

\bibitem{wangPhononMediatedLongRangeAttractive2021}
Yao Wang, Zhuoyu Chen, Tao Shi, Brian Moritz, Zhi-Xun Shen, and Thomas~P.
  Devereaux.
\newblock Phonon-{{Mediated Long-Range Attractive Interaction}} in
  {{One-Dimensional Cuprates}}.
\newblock {\em Physical Review Letters}, 127(19):197003, November 2021.

\bibitem{senechalSpectralWeightHubbard2000a}
D.~S{\'e}n{\'e}chal, D.~Perez, and M.~{Pioro-Ladri{\`e}re}.
\newblock Spectral weight of the hubbard model through cluster perturbation
  theory.
\newblock {\em Physical Review Letters}, 84(3):522--525, January 2000.

\bibitem{senechalCPT2002}
David S\'en\'echal, Danny Perez, and Dany Plouffe.
\newblock Cluster perturbation theory for hubbard models.
\newblock {\em Phys. Rev. B}, 66:075129, Aug 2002.

\bibitem{whiteDensityMatrixFormulation1992}
Steven~R. White.
\newblock Density matrix formulation for quantum renormalization groups.
\newblock {\em Physical Review Letters}, 69(19):2863--2866, November 1992.

\bibitem{whiteDMRG1993}
Steven~R. White.
\newblock Density-matrix algorithms for quantum renormalization groups.
\newblock {\em Phys. Rev. B}, 48:10345--10356, Oct 1993.

\bibitem{vidalEfficientSimulationOneDimensional2004}
Guifr{\'e} Vidal.
\newblock Efficient {{Simulation}} of {{One-Dimensional Quantum Many-Body
  Systems}}.
\newblock {\em Physical Review Letters}, 93(4):040502, July 2004.

\bibitem{whiteRealTimeEvolutionUsing2004}
Steven~R. White and Adrian~E. Feiguin.
\newblock Real-{{Time Evolution Using}} the {{Density Matrix Renormalization
  Group}}.
\newblock {\em Physical Review Letters}, 93(7):076401, August 2004.

\bibitem{paeckelTimeevolutionMethodsMatrixproduct2019}
Sebastian Paeckel, Thomas K{\"o}hler, Andreas Swoboda, Salvatore~R. Manmana,
  Ulrich Schollw{\"o}ck, and Claudius Hubig.
\newblock Time-evolution methods for matrix-product states.
\newblock {\em Annals of Physics}, 411:167998, December 2019.

\bibitem{zhangDensityMatrixApproach1998a}
Chunli Zhang, Eric Jeckelmann, and Steven~R. White.
\newblock Density {{Matrix Approach}} to {{Local Hilbert Space Reduction}}.
\newblock {\em Physical Review Letters}, 80(12):2661--2664, March 1998.

\bibitem{brocktMatrixproductstateMethodDynamical2015a}
C.~Brockt, F.~Dorfner, L.~Vidmar, F.~{Heidrich-Meisner}, and E.~Jeckelmann.
\newblock Matrix-product-state method with a dynamical local basis optimization
  for bosonic systems out of equilibrium.
\newblock {\em Physical Review B}, 92(24):241106, December 2015.

\bibitem{jeckelmannDynamicalDensitymatrixRenormalizationgroup2002}
Eric Jeckelmann.
\newblock Dynamical density-matrix renormalization-group method.
\newblock {\em Physical Review B}, 66(4):045114, July 2002.

\bibitem{benthien1dHubbardDDMRG2004}
H.~Benthien, F.~Gebhard, and E.~Jeckelmann.
\newblock Spectral function of the one-dimensional hubbard model away from half
  filling.
\newblock {\em Phys. Rev. Lett.}, 92:256401, Jun 2004.

\bibitem{essler1dHubbard2005}
Fabian H.~L. Essler, Holger Frahm, Frank Göhmann, Andreas Klümper, and
  Vladimir~E. Korepin.
\newblock {\em The One-Dimensional Hubbard Model}.
\newblock Cambridge University Press, 2005.

\bibitem{kohnoSpectralPropertiesMott2010}
Masanori Kohno.
\newblock Spectral {{Properties}} near the {{Mott Transition}} in the
  {{One-Dimensional Hubbard Model}}.
\newblock {\em Physical Review Letters}, 105(10):106402, August 2010.

\bibitem{liParticleholeAsymmetryDynamical2021}
Shaozhi Li, Alberto Nocera, Umesh Kumar, and Steven Johnston.
\newblock Particle-hole asymmetry in the dynamical spin and charge responses of
  corner-shared {{1D}} cuprates.
\newblock {\em Communications Physics}, 4(1):1--12, September 2021.

\bibitem{giamarchiQuantumPhysicsOne2003}
Thierry Giamarchi.
\newblock {\em Quantum {{Physics}} in {{One Dimension}}}.
\newblock International {{Series}} of {{Monographs}} on {{Physics}}. {Oxford
  University Press}, {Oxford}, 2003.

\bibitem{pengEnhancedSuperconductivityNearneighbor2022}
Cheng Peng, Yao Wang, Jiajia Wen, Young Lee, Thomas Devereaux, and Hong-Chen
  Jiang.
\newblock Enhanced superconductivity by near-neighbor attraction in the doped
  {{Hubbard}} model, June 2022.

\end{thebibliography}

%%TC:ignore
\appendix

\setcounter{equation}{0}
\setcounter{figure}{0}
\setcounter{table}{0}
\setcounter{page}{1}
\makeatletter
\renewcommand{\theequation}{S\arabic{equation}}
\renewcommand{\thefigure}{S\arabic{figure}}

\section{Supplementary Material}

\subsection{Time-evolving Block Decimation}
We use the time-evolving block decimation (TEBD) scheme, which was invented by Vidal~\cite{vidalEfficientSimulationOneDimensional2004} and later incorporated into the DMRG algorithm by White~\cite{whiteRealTimeEvolutionUsing2004} for time evolution. TEBD utilizes a Trotter-Suzuki decomposition of the time evolution operators; and we consider the second-order TEBD (TEBD2) scheme, which uses the decomposition 
\begin{equation}
    e^{-i H \delta t} \!=\! \prod_{i=0}^{L-2}\!\!e^{-i h_{i,i+1}\delta t/2}\prod_{i=L-2}^{0}\!\!e^{-i h_{i,i+1}\delta t/2} + O(\delta t^3).
    \label{eq:ts3}
\end{equation}
Here, we consider a Hamiltonian that contains only nearest-neighbor couplings and $h_{i,i+1}$ contains terms involving only sites $i$ and $i+1$ along a 1D chain. 
The time evolution operator can be applied to the wave function during a DMRG sweep, replacing the ground state solving step by applying the gate $e^{-i h_{i,i+1}\delta t/2}$ when sites $i$ and $i+1$ are at the center. In this way, we avoid numerical errors due to truncation when site $i$ or $i+1$ is in the system or environment block. 
Specifically, we can apply gates $e^{-i h_{0,1}\delta t/2}$, $e^{-i h_{1,2}\delta t/2}$, $\dots$, $e^{-i h_{L-2,L-1}\delta t/2}$ in the left-to-right sweep, then reverse sweep direction, and apply all the reverse gates in the right-to-left sweep. Thus all gates for one time step can be applied by a complete left-to-right and right-to-left sweep~\cite{whiteRealTimeEvolutionUsing2004, paeckelTimeevolutionMethodsMatrixproduct2019}.

\subsection{Local Basis Optimization}
The unbounded phonon Hilbert space on each site presents a challenge for wave function based numerical techniques. 
It is usually inefficient to naively truncate the Hilbert space, keeping only the first $N$ bare phonon basis on each site ($\left|0\right>$, $\left|1\right>$, ..., $\left|N-1\right>$), especially when the {\it e-ph} coupling is strong and many bare phonons are needed for convergence. This becomes prohibitive for techniques like exact diagonalization (ED) and also may make DMRG simulations difficult, if not unfeasible. 
One method to solve this problem is to perform a local basis optimization (LBO), truncating the local phonon Hilbert space to a few optimal basis~\cite{zhangDensityMatrixApproach1998a}, similar to truncation of the system and environment Hilbert space blocks in traditional DMRG ground state calculations. 
This approach works very well, often with only $2$ to $3$ optimal phonon basis elements can provide good ground state convergence in the Holstein model~\cite{zhangDensityMatrixApproach1998a}.

LBO has been extended for time evolution, a dynamical LBO, where phonon basis on each site is optimized in a position- and time-dependent manner~\cite{brocktMatrixproductstateMethodDynamical2015a}. Fig.~\ref{pic:diagram}(b) illustrates how to perform dynamical LBO, where the local Hilbert space of dimension $D$ is optimally truncated to $d \ll D$. During time evolution, We first enlarge the Hilbert space of each of the two center sites to dimension $D$, then apply the Trotter gate in this enlarged Hilbert space to reduce errors due to truncation. Finally, we truncate the local Hilbert space back to dimension $d$, which significantly reduces the numerical cost for truncating the system or environment block~\cite{brocktMatrixproductstateMethodDynamical2015a}. In our calculations, both ground state LBO and dynamical LBO for time evolution provide reasonable convergence for the Hubbard-extended Holstein model.

\subsection{Lesser Green's function and Fourier Transform}
Using time evolution, we can calculate the lesser Green's function
$\mathcal{G}^{<}_{mn\sigma}(t) = i\left<\hat{c}^\dagger_{m\sigma}(t) \hat{c}_{n\sigma}(0)\right>$.
To do so, we need to time evolve both the ground state $\left|G(t)\right> = e^{-iHt}\left|G\right>$ and the removal state $\left|R_{n\sigma}(t)\right> = e^{-iHt}\hat{c}_{n\sigma}\left|G\right>$, such that $\mathcal{G}^{<}_{mn\sigma}(t) = i\left<G(t)\right|\hat{c}^\dagger_{m\sigma}\left|R_{n\sigma}(t)\right>$. 
The single-particle removal spectra, which can be compared to ARPES spectra, is obtained by a Fourier transform of the lesser Green's function
\begin{equation}
    \mathcal{A}^{-}(k,\omega) = \int^\infty_{-\infty} \frac{dt ~e^{i\omega t}}{2\pi i} \sum_{mn\sigma} \frac{e^{-ik(r_n - r_m)}}{L^2} \mathcal{G}^{<}_{mn\sigma}(t),
\end{equation}
where $k$ represents momentum along the chain, $\omega$ is frequency, and $L$ is the length of the chain.
One typically fixes the position index $n$ to the center of the chain ($n=L/2-1$ or $n = L/2$); correspondingly, the summation runs only over index $m$ with a normalization factor $1/L$ rather than $1/L^2$. To ensure reflection symmetry for chains with an even number of sites, we average over the spectra obtained from $\mathcal{G}^{<}_{m,L/2-1,\sigma}$ and $\mathcal{G}^{<}_{m,L/2,\sigma}$. 
%where $\mathcal{G}_{m,L/2,\sigma}$ is obtained from $\mathcal{G}_{m,L/2-1,\sigma}$ by reversing the chain index. 
We evolve the system from time 0 to time T, before the excitation propagates to the boundary of the chain. To regularize the Fourier transform due to the finite cutoff time, we use a window function $\mathcal{W}(t)$, where $\mathcal{W}(T) \sim 0$, which broadens the spectra and acts as a frequency resolution convolution 
$\mathcal{A}(k,\omega) = \mathcal{A}_{-}(k,\omega) * \mathcal{F}[\mathcal{W}(t)]$. 
We use a Gaussian window function with frequency domain standard deviation $\sigma_{\omega}$. We also convolve the spectra in momentum using a Gaussian filter with a standard deviation of one momentum spacing $\sigma_{k} = 2\pi/L$, which removes high frequency noise and smooths the spectra.

\subsection{Convergence}
We keep $m=800$ states during time evolution which produces a truncation error below $1\times 10^{-6}$ for the Hubbard and extended Hubbard models, and below $1\times 10^{-5}$ for the Hubbard-extended Holstein model. The time step is fixed at $\delta t = 0.025 t_h^{-1}$, and we evolve the system up to $T=20 t_h^{-1}$. For dynamical LBO, we keep $20$ bare phonon basis ($D=80$ in Fig.~\ref{pic:diagram}(b)) and truncate to a basis of $3$ optimal phonons ($d=12$ in Fig.~\ref{pic:diagram}(b)) on every site. This results in a phonon truncation error below $1\times 10^{-4}$. 

\begin{figure}[htpb!]
    \begin{center}
        \includegraphics[width=\picWidth]{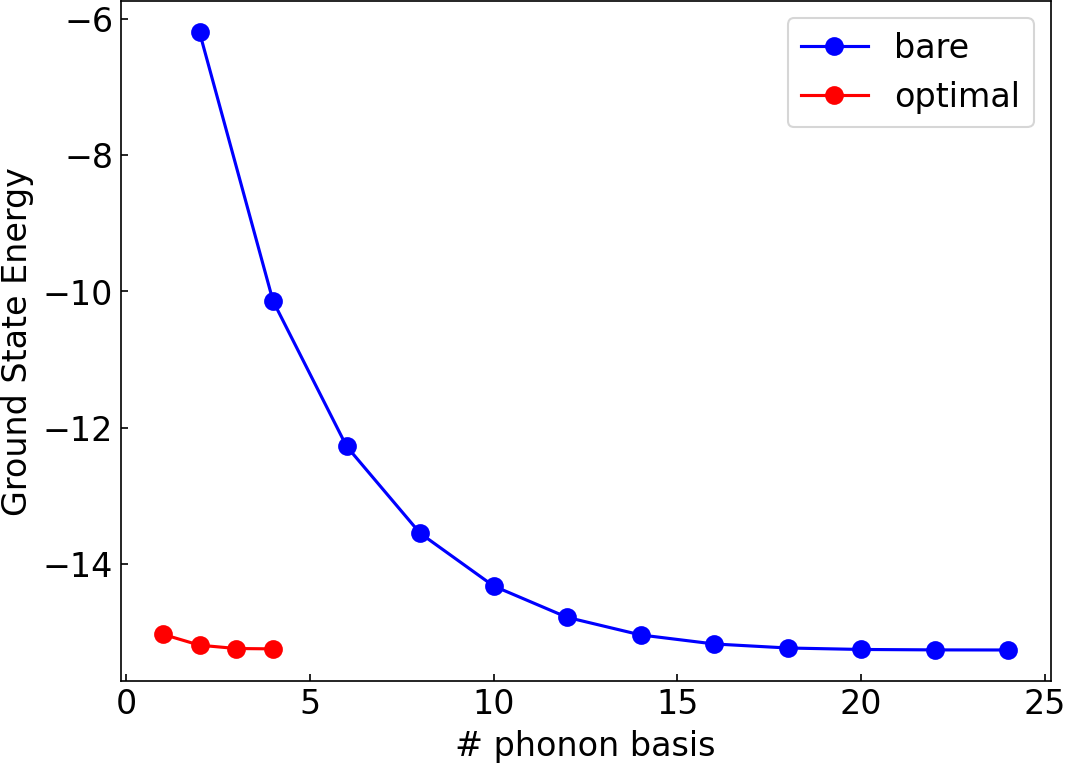} 
    \end{center}
    \caption{
        Ground state energy convergence of LBO on an 8-site chain at half filling. The blue curve corresponds to using bare phonon basis without LBO and we need about 20 bare phonon basis to converge the ground state. The red curve is LBO with 20 bare phonon basis and the ground state is well converged with just 3 optimal phonon basis.
    }
    \label{pic:lbo_convergence}
\end{figure}

\begin{figure}[htpb!]
    \begin{center}
        \includegraphics[width=\picWidth]{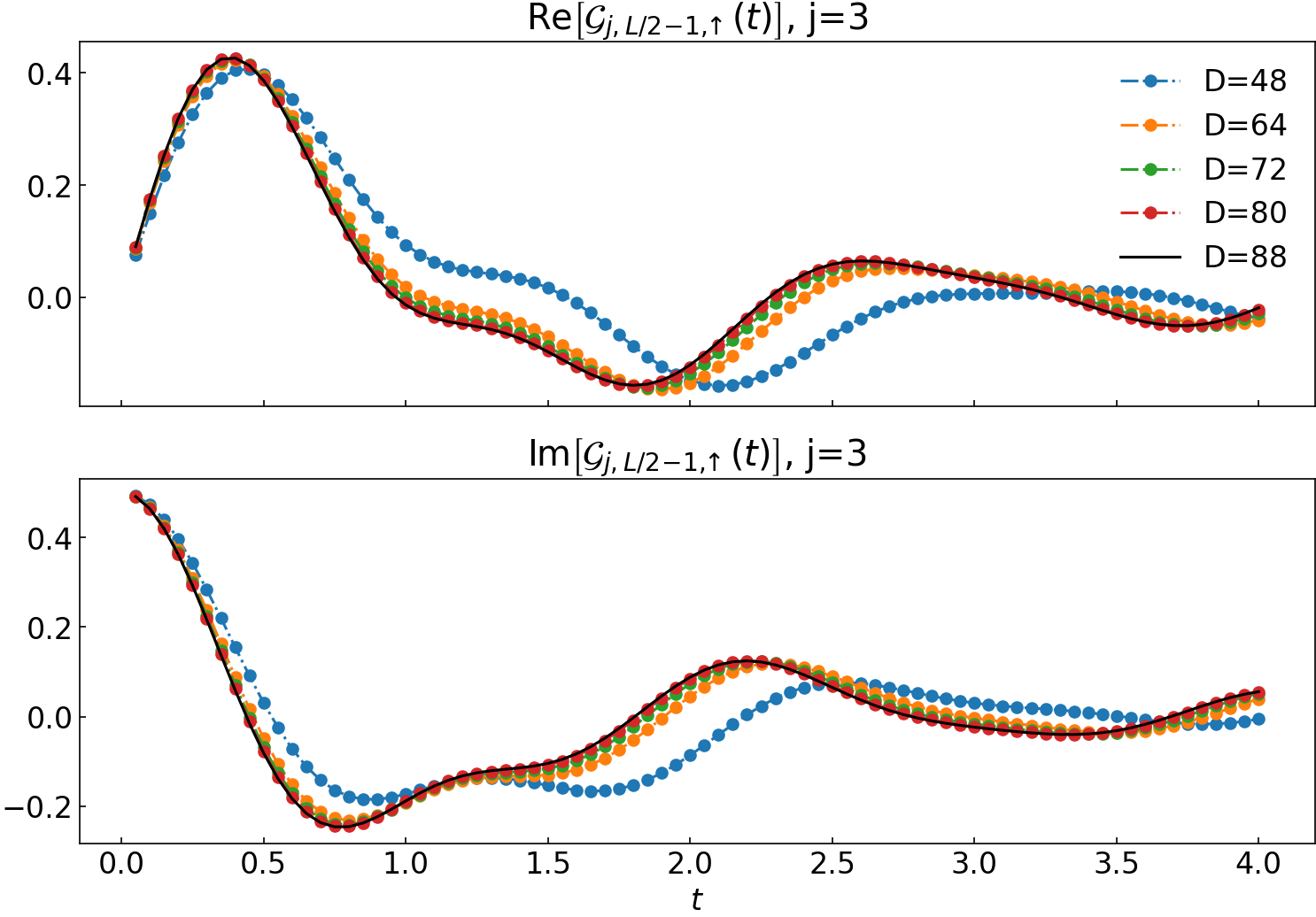} 
    \end{center}
    \caption{
        Time evolution convergence of the lesser Green's function with respect to the number of bare local basis dimension $D$ on an 8-site chain at half filling without dynamical LBO. The time step is $\delta t = 0.05$ here. The lesser Green's function is well converged with $D=80$ (20 bare phonon basis). 
    }
    \label{pic:time_nBare_convergence}
\end{figure}

\begin{figure}[htpb!]
    \begin{center}
        \includegraphics[width=\picWidth]{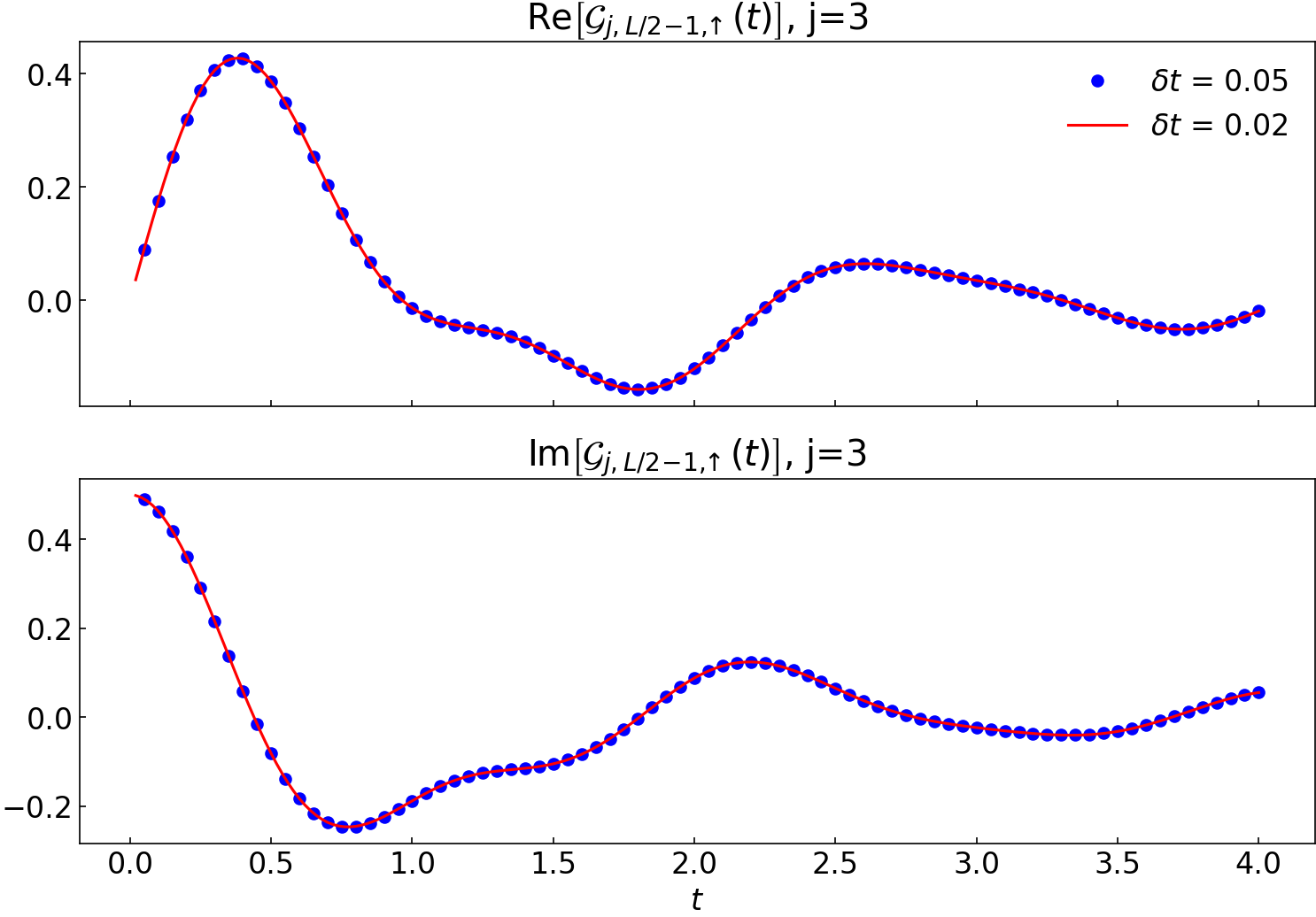} 
    \end{center}
    \caption{
        Time evolution convergence of the lesser Green's function with respect to time step on an 8-site chain at half filling without dynamical LBO. We use $D=80$ here and we see that time evolution with $\delta t = 0.05$ is well converged. 
    }
    \label{pic:timeStep_convergence}
\end{figure}

\begin{figure}[htpb!]
    \begin{center}
        \includegraphics[width=\picWidth]{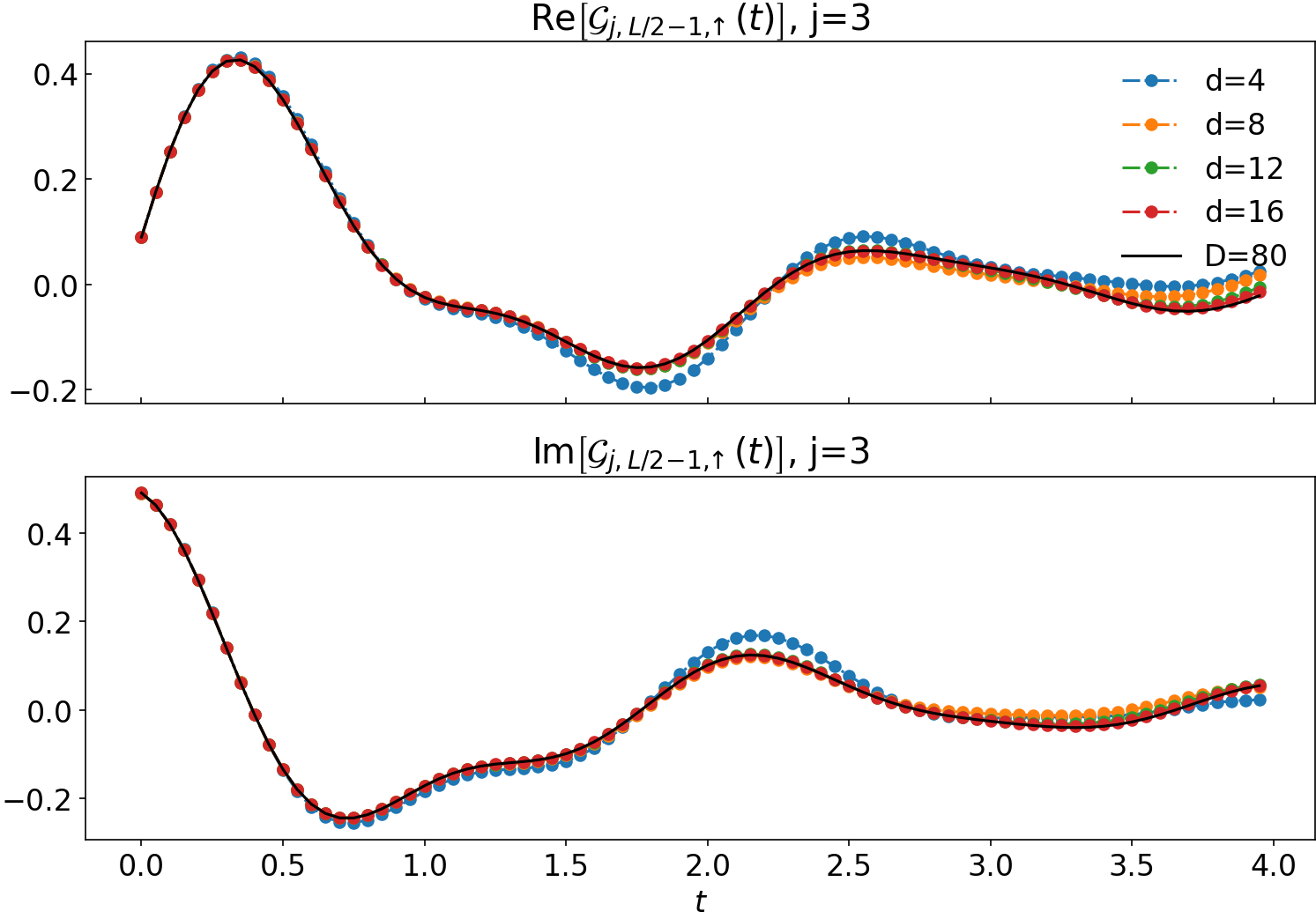} 
    \end{center}
    \caption{
        Time evolution convergence of the lesser Green's function with respect to optimal basis number with dynamical LBO. We fix $D=80$ and $\delta t=0.05$ and we see $d=12$ is enough for good convergence. 
    }
    \label{pic:dlbo_convergence}
\end{figure}

Time evolution convergence with respect to both $\delta t$ and $m$ has been checked on the $80$-site chain for the Hubbard model. Adding phonons makes the calculations quiet expensive and convergence with respect to the local bare basis dimension $D$ and optimal basis dimension $d$ have been checked on an 8-site chain, where many more bare phonons can be kept and time evolution without dynamical LBO can be carried out for benchmark. For $D=80$ and $d=12$, both the ground state energy (see Fig.~\ref{pic:lbo_convergence}) and the lesser Green's function (see Fig.~\ref{pic:time_nBare_convergence},  \ref{pic:timeStep_convergence} and \ref{pic:dlbo_convergence}) converge well on the short chain for the Hubbard-extended Holstein model, and time evolution on the 80-site chain can be completed for a reasonable computational cost. 
%Since the computation time grows quickly with respect to $m, D$ and $d$ \cite{brocktMatrixproductstateMethodDynamical2015a}, GPU accelerated tDMRG algorithm will be helpful to further improve accuracy. 

In the ground state correlation functions calculation, we keep up to $m=1000$ states, which results in a truncation error ranging from $6\times 10^{-9}$ to $3\times 10^{-7}$, depending on the model and {\it e-ph} coupling strength. We truncate a bare phonon basis of up to 40 ($D=160$) to an optimal phonon basis of up to 4 ($d=16$). This results in a phonon basis truncation error ranging from  $2\times 10^{-8}$ to $3\times 10^{-7}$, depending on the {\it e-ph} coupling strength. 

\begin{figure*}[htpb!]
    \begin{center}
        \includegraphics[width=\widePicWidth]{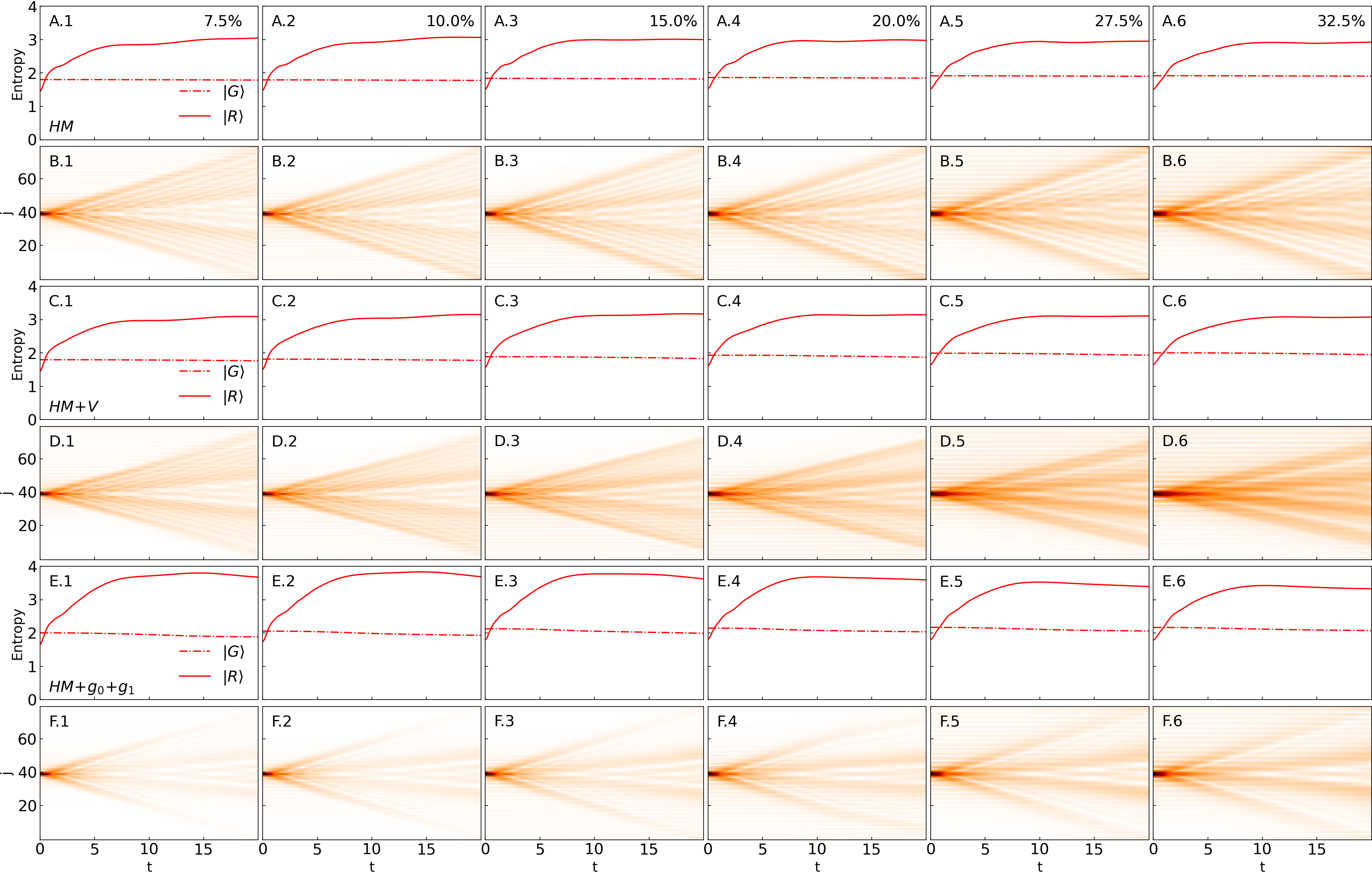} 
    \end{center}
    \caption{
        Raw data of entropy growth during time evolution and lesser Green's function of different models at different dopings.
    }
    \label{pic:green}
\end{figure*}

\begin{figure*}[htpb!]
    \begin{center}
        \includegraphics[width=\widePicWidth]{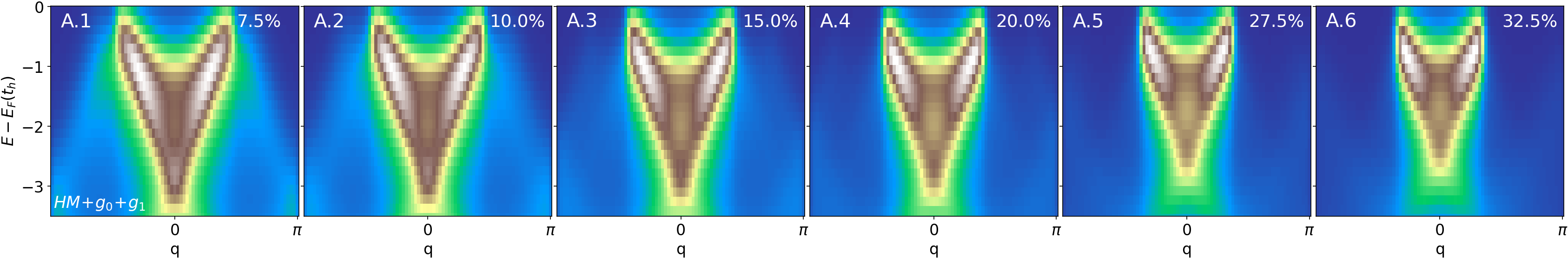} 
    \end{center}
    \caption{
        Single particle spectral function for the Hubbard-extended Holstein model with larger energy broadening: $0.3t_h$ Lorentzian broadening plus $0.2t_h$ additional Gaussian broadening, similar to the broadening used in Ref.\cite{chenAnomalouslyStrongNearneighbor2021}.
    }
    \label{pic:akw_phonon_broad}
\end{figure*}

\begin{figure}[htpb!]
    \begin{center}
        \includegraphics[width=\picWidth]{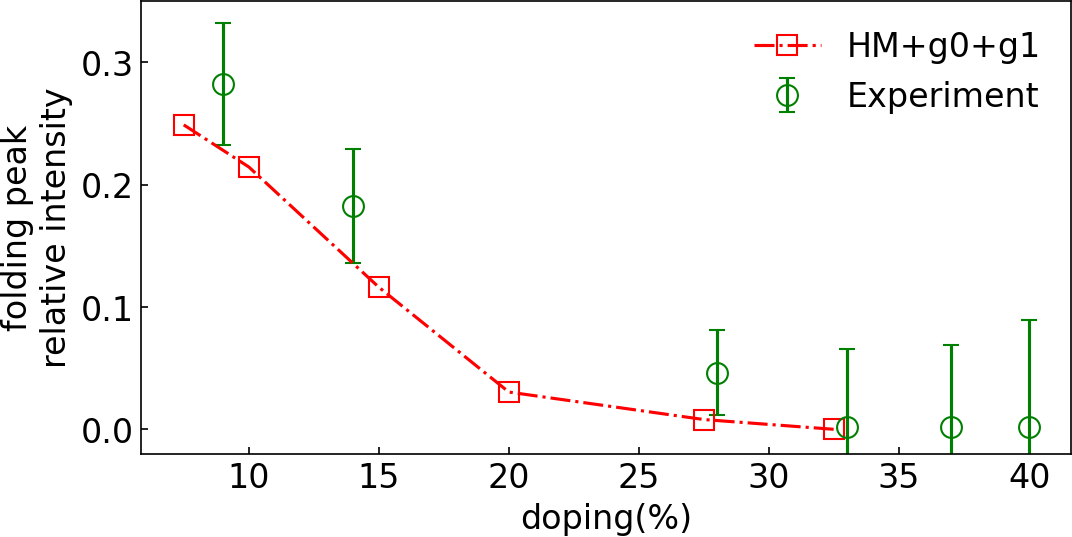} 
    \end{center}
    \caption{
         Holon folding peak intensity relative to the main holon peak as a function of doping. The red square represents data from simulation of Hubbard-extended Holstein model. The intensity is extracted by fitting MDC cut with sum of Gaussian peaks. The MDC cut is chosen to be $\sim t_h$ above the holon bottom and is extracted from Fig.~\ref{pic:akw_phonon_broad}. The green circle represents experimental data from Ref.\cite{chenAnomalouslyStrongNearneighbor2021}.
    }
    \label{pic:phonon_hf_intensity}
\end{figure}

%%TC:endignore
\end{document}